\documentclass[nofootinbib,showpacs, aps,prb,twocolumn,preprintnumbers ,amsmath, amssymb, superscriptaddress, aps]{revtex4-2}
\usepackage{lineno}

\usepackage{color}
\usepackage{amsmath,amssymb}
\usepackage[mathscr]{euscript}
\usepackage{pifont}
\usepackage{amssymb}  
\usepackage{bbold}
\usepackage{float}
\usepackage{graphicx} 
\usepackage{dcolumn}  
\usepackage{bm}       
\usepackage[colorlinks]{hyperref}

\usepackage{dcolumn}
\newcolumntype{d}[1]{D{.}{.}{#1}}
\usepackage{siunitx}

\usepackage{stackengine}

\def\g#1{{g^{(#1)}}}

\def\kB{{k_{\rm B}}}
\def\kBT{{k_{\rm B}T}}
\def\sp3{{sp^3}}
\def\avg#1{{\langle #1\rangle }}
\def\rmd{{\rm d}}
\def\Tr{{\rm Tr}}
\def\kB{{k_{\rm B}}}
\def\kBT{{k_{\rm B}T}}
\def\bp{{\bf p}}
\def\br{{\bf r}}
\def\cH{{\mathcal H}}
\def\cN{{\mathcal N}}
\def\cV{{\mathcal V}}

\newcommand{\Lim}[1]{\raisebox{0.5ex}{\scalebox{0.8}{$\displaystyle \lim_{#1}\;$}}}

\begin{document}

\title{Reversible non-equilibrium phase transformation in amorphous germanium}

\author{Yang Huang}
\affiliation{
  University of Science and Technology of China, Hefei 230026, China
}
\affiliation{
  Suzhou Institute for Advanced Research, University of Science and Technology of China, Suzhou 2
15213, China
}

\author{Marek Mihalkovič}
\affiliation{Institute of Physics, Slovak Academy of Sciences, SK-84511 Bratislava, Slovakia}

\author{Michael Widom}
\affiliation{Carnegie Mellon University, Department of Physics, Pittsburgh, 15213, PA, USA}

\date{\today }

\begin{abstract}
First principles molecular dynamics simulations of germanium reveal a reversible liquid-glass transition below the equilibrium melting point with a wide hysteresis loop. Direct calculation of the liquid and amorphous state free energies, enabled by absolute entropy calculations, verify that the transition is first order in character between two metastable phases. These results lend credence to models of explosive recrystallization from amorphous Ge thin films, through a metastable liquid state and finally reaching the low temperature crystalline structure.
\end{abstract}

\maketitle 


{\em Introduction---}Elements from group IVB of the periodic table such as carbon, silicon, germanium, and tin, form low density four-coordinated crystal structures featuring tetrahedral $\sp3$-type covalent bonds. Some molecular crystals, notably solid water (ice) mimic tetrahedral bonding~\cite{PaulingIce}. When these structures melt, the low density, open structure is destroyed, resulting in higher density and more highly coordinated, but less ordered, structures. The change in structure when ice melts to water has been linked with the anomalous negative thermal expansion of water close to its melting point. This even motivated the proposal that a critical point might exist in supercooled water, below which a high density liquid (HDL) and a low density liquid (LDL) might coexist~\cite{Stanley,Water}. Through the analogy between tetrahedral bonding of water and $\sp3$-bonded elements, similar liquid-liquid phase transitions have been proposed in carbon~\cite{LLPT-C,LiquidC}, silicon~\cite{SastryAngell,Ganesh09}, germanium~\cite{Bhat}, and tin~\cite{tin}, mainly on the basis of computer simulations.

\begin{figure}
  \includegraphics[width=0.45\textwidth,trim=0cm 7cm 4cm 0cm,clip]{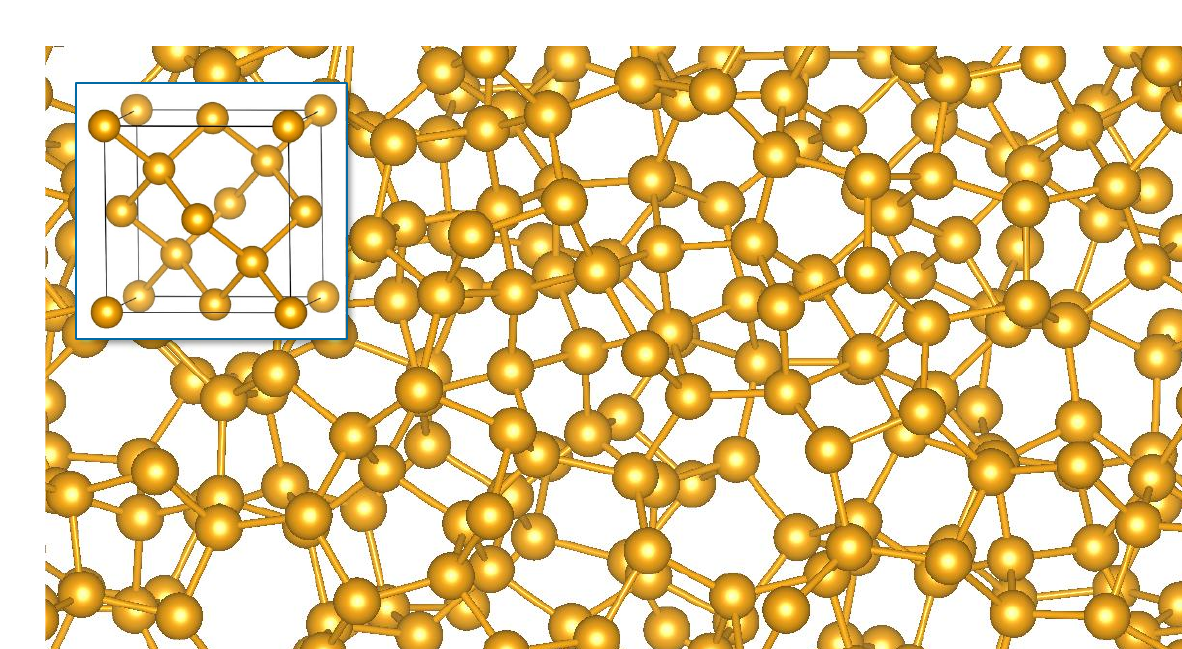}
  \caption{\label{fig:a-Ge} Simulated structure of $\sp3$-bonded amorphous Ge. The glassy structure is almost perfectly four-fold coordinated. Inset: crystalline diamond structure.}
\end{figure}

While the crystalline solid and liquid states are assumed to be equilibrium states (or metastable in the supercooled cases), the group IVB elements, and water, also exist out of equilibrium in glassy, or amorphous, states (see Fig.~\ref{fig:a-Ge}). Additionally, they exhibit {\em two} such states, high density amorphous (HDA) and low density amorphous (LDA).  The LDA state forms at ordinary pressures, while large applied pressure creates HDA. These states have been confirmed experimentally in water~\cite{Mishima}, silicon~\cite{McMillan} and germanium~\cite{HDA-Ge}.

Synthesis of amorphous Ge requires extreme quench rates and this is often achieved through physical vapor deposition of amorphous thin films~\cite{Tauc1966}. A glass transition upon cooling the bulk liquid is not observed experimentally owing to the ease of crystallization. Recrystallization of amorphous thin films can be ``explosive''~\cite{Takamori1972}, proceeding at rates above 200 cm/sec~\cite{Chapman1980}. Recrystallization is believed to be mediated by a thin interface of liquid germanium at the amorphous to crystal interface~\cite{Leamy1981,Nikolova2014}. Thermodynamic analysis suggests that the liquid is metastable and forms well below the crystalline melting temperature of 1210K~\cite{Bagley1979}.

\begin{figure}
  \includegraphics[width=0.45\textwidth]{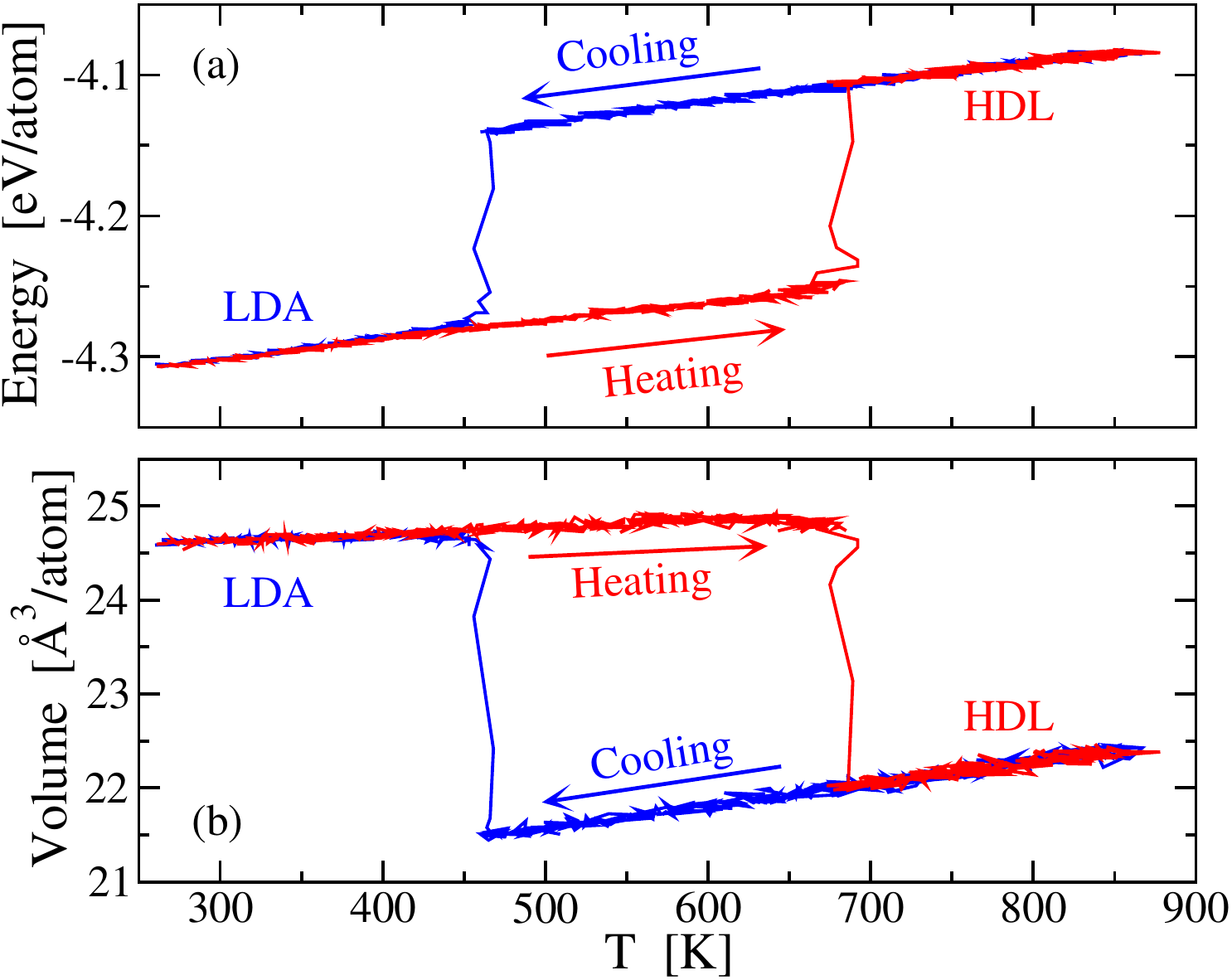}
  \caption{\label{fig:loop} Hysteresis loops in (a) energy and (b) volume for 4096 Ge atoms at rates of 10$^{10}$ K/sec. Values are plotted at intervals of 50 ps.}
\end{figure}

Here we focus on temperature driven phase transformations of germanium at low pressure. The low temperature crystalline state takes the Pearson type cF8 diamond structure, $\alpha$-Ge, with perfect tetrahedral coordination. Electronic structure calculations reveal a narrow band gap, confirming its semiconducting character.  Applying {\em ab-initio} molecular dynamics simulations (AIMD) in the Gibbs (NPT) ensemble, we can heat up and eventually melt the crystal structure, at which point its density jumps and it enters a metallic high density liquid (HDL) state. Upon cooling we are unable to restore the crystalline state owing to our high quench rate. Instead, it transforms to a low density four-coordinated $\sp3$-bonded amorphous (LDA) solid phase. Our simulation resembles the pioneering application of first principles calculations to a quench of liquid Ge by Kresse and Hafner~\cite{KresseHafnerGe}, but with a sample size 64 times larger and a quench rate 10$^3$ times lower. Subsequent heating and cooling reveals a reversible but strongly hysteretic transition, as shown in Fig.~\ref{fig:loop}, suggestive of a first order phase transition between the low temperature semiconducting LDA state and the high temperature metallic liquid HDL state.

Because our hysteresis loop is so wide, we cannot identify a precise transition temperature. Instead, we combine our simulated HDL-LDA enthalpy difference $\Delta H$ with calculations of the entropy difference $\Delta S$ in order to identify the temperature $T=\Delta H/\Delta S$ at which the Gibbs free energies of the two phases cross. While enthalpies are calculated directly from the total energies calculated during our simulations, calculation of the entropies requires distinct methods of analysis.

Experimentally, entropy is determined by relating the state of interest to some reference state of known entropy, $S_{\rm ref}$. The change in entropy between the two states can be evaluated by integration of the heat capacity divided by temperature.
\begin{equation}
  \label{eq:S_thermo}
S(P,T) = S_{\rm ref}+\int_{T_{\rm ref}}^T \rmd T' ~C_p(T')/T'.
\end{equation}
The reference state can be taken as the limit of low temperature, $T\to 0$K, and the convention is to require $S(T=0)=0$ in accordance with the third law of thermodynamics. Unfortunately, it is often difficult to impose this condition accurately, especially when passing through phase transitions. Since the entropy is a function of thermodynamic state, there must exist a means to determine it directly from the state without the need for thermodynamic integration. The statistical entropy~\cite{Gibbs,Shannon1948}
\begin{equation}
  \label{eq:S_stat}
  S = -\kB\Tr(\rho\ln{\rho})
\end{equation}
automatically yields values that are consistent with the third law and requires no artificial reference value. Here $\rho$ is the many-body density function that gives the probability as a function of all atomic coordinates in the classical case, while in the quantum case the density operator is required.

In the following, we apply an expansion of the statistical entropy in a series of few-body correlation functions to evaluate the entropy of the high density liquid state. We then evaluate the entropy of the low density amorphous state by adding its quantum vibrational entropy to the entropy of the underlying covalent bonding topology. The resulting free energies cross at $T=647$K,which lies inside our hysteresis loop, confirming the anticipated first order nonequilibrium phase transition.


{\em High density liquid---}As we wish to model the phase transition at fixed pressure we require the Gibbs free energy $G(N,P,T)=H-TS$, where $H=E+PV$ is the enthalpy. We choose pressure $P=0$, so the energy and enthalpy are numerically equal, provided the volume $V$ matches the equilibrium volume. Hence the Gibbs free energy equals the Helmholtz free energy $F(N,V,T)=E-TS$. The energy was obtained by averaging the VASP total energies $E0$ for the duration of our data collection runs. As these are only potential energies, we add the kinetic energy $K=3\kB/2$, as shown in Table~\ref{tab:F_HDL}.

\begin{table}[b]
  \begin{tabular}{c|rrrrr}
    T          &      500 &    550 &    600 &    650 &    700\\
    \hline
    $\avg{K}$  &    0.065 &  0.071 &  0.078 &  0.084 &  0.090\\
    $\avg{E_0}$&   -4.128 & -4.121 & -4.116 & -4.108 & -4.100\\
    $\avg{H}$  &   -4.064 & -4.050 & -4.039 & -4.024 & -4.010\\
    \hline
    $S_1$      &   11.737 & 11.884 & 12.019 & 12.147 & 12.266\\
    $S_2$      &   -1.669 & -1.598 & -1.535 & -1.471 & -1.407\\
    $S_3$      &   -0.404 & -0.351 & -0.304 & -0.262 & -0.231\\
    $S$        &    9.654 &  9.926 & 10.173 & 10.408 & 10.615\\
    \hline
    $G$        &   -4.481 & -4.522 & -4.567 & -4.610 & -4.653
  \end{tabular}
  \caption{\label{tab:F_HDL} Energy terms (in units of eV/atom), entropy terms (in units of $\kB$/atom), and free energies of the high density liquid.  The Gibbs free energy $G=H-TS+F_e$ with $S=S_1+S_2+S_3$.}
\end{table}

The entropy of a liquid can be expanded in a series of few-body terms~\cite{Nettleton1958,Raveche1971a,Evans1989}
\begin{equation}
  \label{eq:S_series}
S = S_1 + S_2 + S_3 + \cdots
\end{equation}
where the $n^{\rm th}$ term requires integrating an $n$-body correlation function. The one-body term depends only on the mean density $\rho$,
\begin{equation}
  \label{eq:S1}
  S_1/\kB = \frac{3}{2} - \ln{\rho\Lambda^3},
\end{equation}
where $\Lambda=\sqrt{h^2/2\pi m k_BT}$ is the thermal De~Broglie wavelength. The first term in $S_1$ represents the kinetic entropy that arises from the Gaussian (Maxwell-Boltzmann) distribution of momenta. The second term is the positional entropy of a single particle localized within one $\Lambda^3$ subvolume within its average volume of $1/\rho$.

\begin{figure}
  \includegraphics[width=0.45\textwidth]{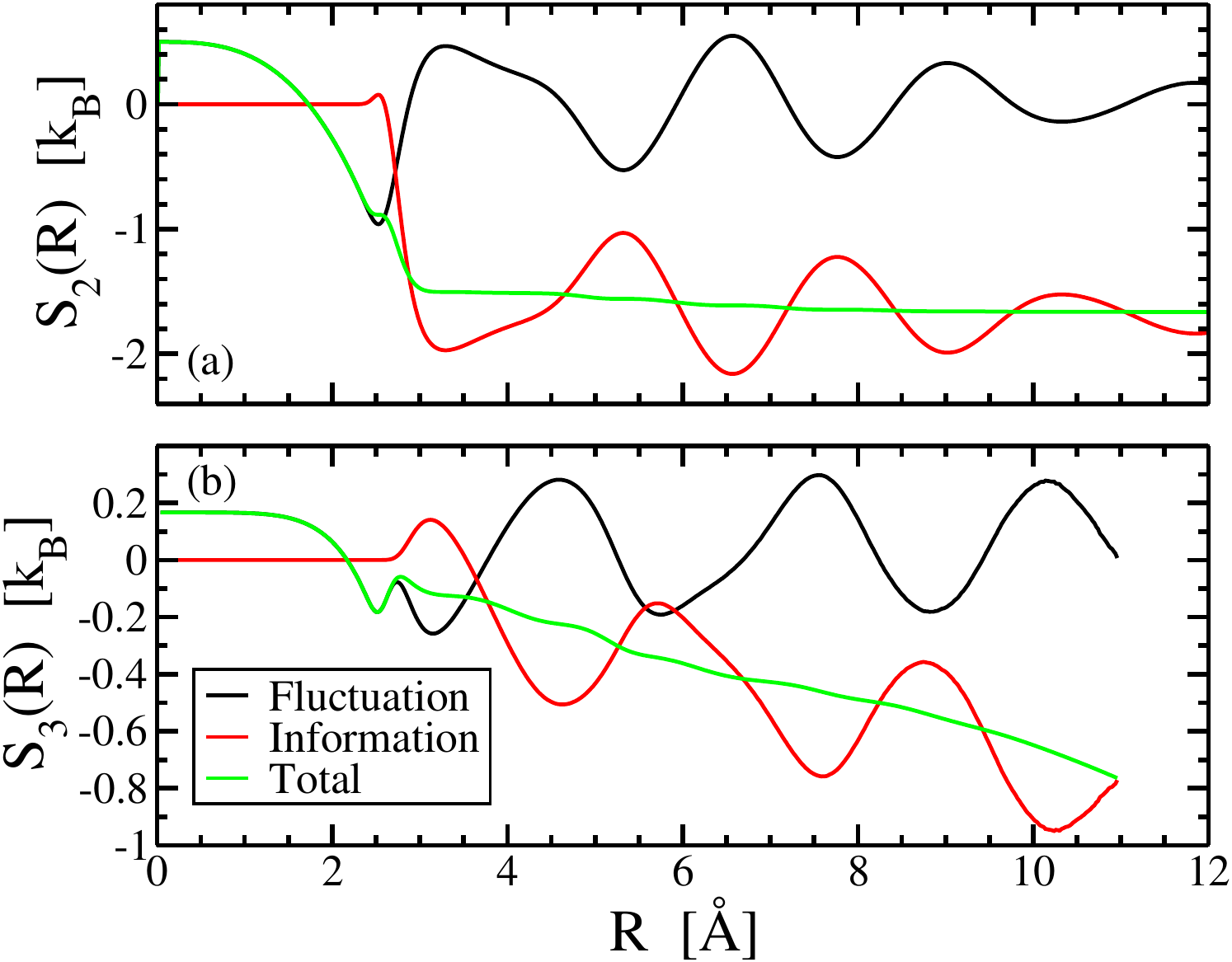}
  \caption{\label{fig:integrals} Integrals in equations~\eqref{eq:S2} and~\eqref{eq:S3} evaluated up to radius $R$. Figures show the fluctuation and information terms~\cite{AlEntropy,ThreeBody} separately, and their sums.}
\end{figure}

For the HDL metallic liquid we include two- and three-body terms~\cite{InfoEntropy,AlEntropy,ThreeBody}. The radial distribution function $\g{2}(r)$ yields the two-body term~\cite{InfoEntropy,AlEntropy}
\begin{equation}
  \label{eq:S2}
  \begin{split}
    S_2/\kB &= \frac{1}{2} + \frac{1}{2}\rho\int\rmd\br~ [\g{2}(r)-1 \\
            & -\g{2}(r)\ln{\g{2}(r)}].
    \end{split}
\end{equation}
We require the three-body correlation function $\g{3}(r_1,r_2,r_3)$ to obtain~\cite{ThreeBody}
\begin{equation}
  \label{eq:S3}
  \begin{split}
    S_3/\kB &= \frac{1}{6} + \frac{1}{6}\rho^2\int\rmd\br^2~ [\g{3}-\g{2}\g{2}\g{2}\\
    &+(\g{2}-1)^2 -\g{3}\ln{(\g{3}/\g{2}\g{2}\g{2})}].
  \end{split}
\end{equation}
In principle these integrals should extend to infinity, but fortunately their integrands decay rapidly enough that we can obtain adequately converged values up to a range $R$ that is less than half of the system size $L$ for $S_2$, and one quarter of $L$ for $S_3$. Fig.~\ref{fig:integrals} shows examples of this convergence at T=500K (the worst case).

\begin{figure}
  \includegraphics[width=0.45\textwidth]{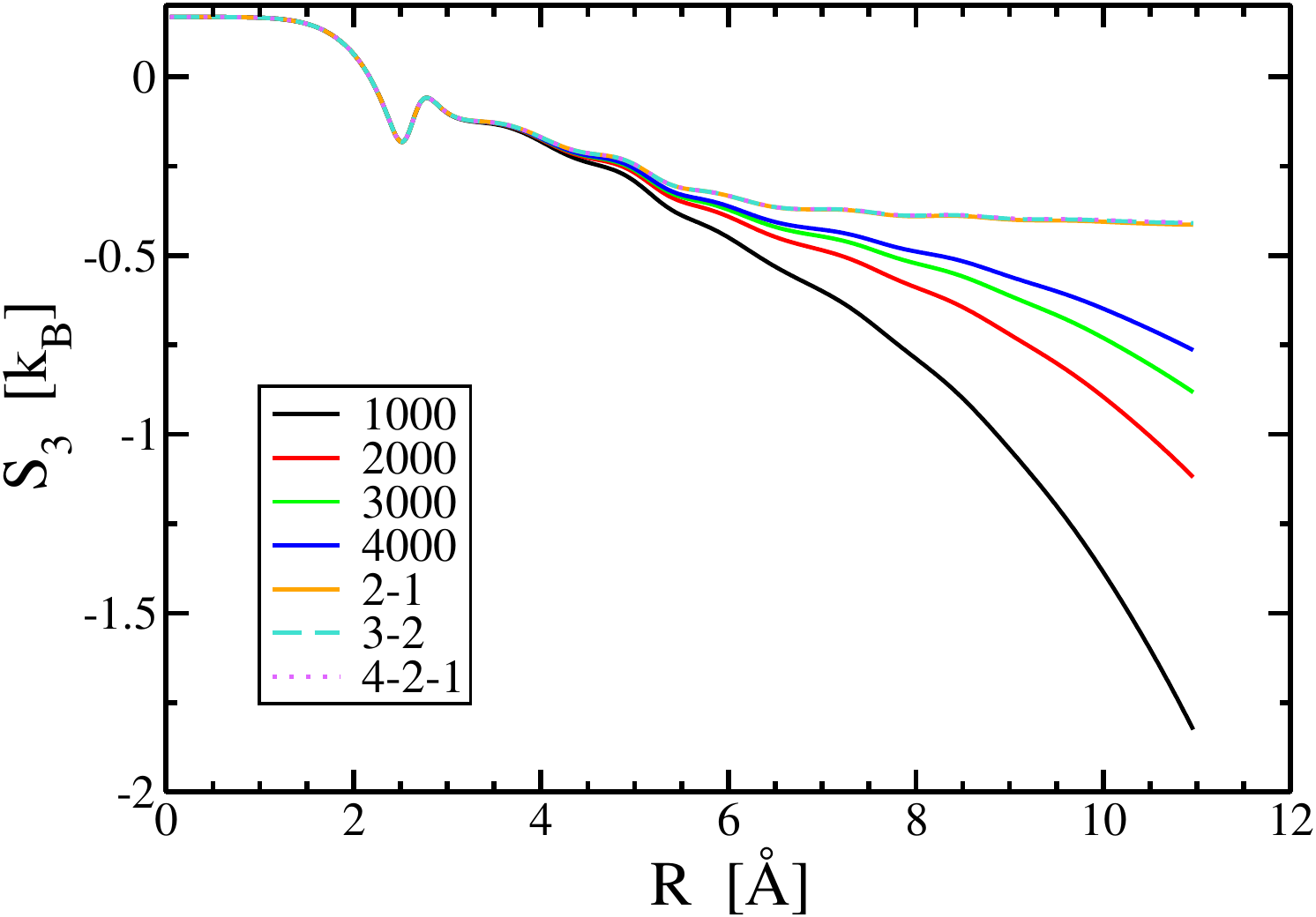}
  \caption{\label{fig:extrap} Plots of $s_t\equiv S_3(R)$ evaluated for data sets of length $t$. First and second-order extrapolations are also shown.}
  \vspace{-0.5cm}
\end{figure}

Notice that the integral $S_3(R)$ does not appear to converge by $R=L/4\approx 12$~\AA. This is a result of insufficient sampling that creates random deviations in $\g{3}$ and, in turn, exaggerates the entropy loss due to false information content~\cite{ThreeBody}. We can correct for this by extrapolating with respect to cumulative run time $t$~\cite{Richardson}. The combination $2S(2t)-S(t)$ cancels out corrections of order $1/t$, and $(8/3)S(4t)-2S(2t)+(1/3)S(t)$ cancels out $1/t^2$ in addition, as we illustrate in Fig. ~\ref{fig:extrap}. Even with the extrapolation there remains a weak downwards slope that we attribute to correlation function errors related to system size. However, the values are sufficiently converged for our needs.

Finally, we include the electronic free energy, $F_e$, which can be calculated from the electronic density of states (electronic densities of states are shown in Fig.~\ref{fig:eDOS} for each structure) and is roughly proportional the value at the Fermi energy~\cite{Chap8}. This small correction ranges from 2-3 meV/atom for HDL but is entirely negligible for the low density amorphous and diamond crystalline structures. On a technical side note, the ML potentials are fitted to the fictitious electronic free energies that VASP defines based upon the applied Fermi level smearing. The differences between these values and the values of $E0$ that are extrapolated to zero smearing turn out to be negligible at the level required for our purposes.


{\em Low density amorphous phase---}As discussed in the introduction, the low density amorphous phase is trapped out of equilibrium in a glassy state with extremely slow dynamics. On the time scale of our simulations the covalent bonding networks hardly evolve. Because the bonds are so strong, the thermal expansion is minimal, and the main effect of temperature is to create small amplitude atomic vibrations. As a result it is an excellent approximation to separate the free energy into a configurational part $F_{\rm conf}=-T S_{\rm conf}$ that describes the bonding network, and a vibrational part $F_{\rm vib}$ that captures the small atomic displacements (see Table~\ref{tab:F_LDA}).

\begin{table}[b]
  \begin{tabular}{c|rrrrr}
        T        &    500 &    550 &    600 &    650 &    700\\
  \hline
  $\avg{K}$      &  0.065 &  0.071 &  0.078 &  0.084 &  0.090\\
  $\avg{E_0}$    & -4.302 & -4.295 & -4.287 & -4.278 & -4.268\\
  $\avg{H}$      & -4.237 & -4.223 & -4.210 & -4.194 & -4.178\\
  \hline
  $S_{\rm conf}$ &  0.822 &  0.828 &  0.837 &  0.847 &  0.891\\
  $S_{\rm vib}$  &  5.758 &  6.039 &  6.294 &  6.533 &  6.761\\
  $S$            &  6.580 &  6.867 &  7.131 &  7.380 &  7.652\\
  \hline
  $G$            & -4.521 & -4.549 & -4.580 & -4.609 & -4.640
  \end{tabular}
  \caption{\label{tab:F_LDA} Entropy terms (in units of $\kB$/atom), energies (units of eV/atom) and free energies of the low density amorphous state. Gibbs free energy $G=H-TS$ with $S=S_{\rm conf}+S_{\rm vib}$.}
\end{table}

Formally, the partition function integrates the Boltzmann factor $\exp{(-\cH(r,p)/\kBT)}$ over all spatial coordinates $r\equiv\{\br_i\}$ and momenta $p\equiv\{\bp_i\}$, where the Hamiltonian includes the kinetic and potential energies $K$ and $V$. The potential energy landscape contains many local minima corresponding to distinct covalent bonding networks $\{\nu\in\cN\}$. Consider a specific network $\nu$ and the set of small deviations in its neighborhood, $\cV(\nu)$. We can approximate the partition function as
\begin{equation}
  \begin{split}
  Z &= \frac{1}{N!}\int_V \frac{\rmd r\rmd p}{h^{3N}} e^{-\cH(r,p)/\kBT}\\
  &\approx \sum_{\nu\in\cN}\int_{\cV(\nu)} \frac{\rmd r\rmd p}{h^{3N}} e^{-\cH(r,p)/\kBT}
\end{split}
\end{equation}
For a given network configuration $\nu$ the integral over small displacements represents the vibrational partition function of that network, allowing us to re-express
\begin{equation}
  Z=\sum_{\nu\in\cN} e^{-F^{(\nu)}_{vib}/\kBT}. 
\end{equation}
In the harmonic approximation, the vibrational free energy of network $\nu$ can be calculated from its vibrational density of states through
\begin{equation}
  \label{eq:F_vib}
  F^{(\nu)}_{\rm vib} = \kBT \int \rmd\omega~ D_\nu(\omega)~\ln[2\sinh{(\hbar\omega/2\kBT)}].
\end{equation}
Because the quantum state determines the probability distribution of both position and momentum, there is no separate kinetic entropy term as was employed in the high density liquid case.

The vibrational free energy depends only weakly on the network configuration $\nu$, so we can further approximate the partition function with the partition function of any randomly drawn configuration multiplied by the number of network configurations $|\cN|$. The free energy becomes
\begin{equation}
  F=-\kBT\ln{Z}=F_{\rm vib}-T S_{\rm conf}
\end{equation}
where the configurational entropy
\begin{equation}
  S_{\rm conf}=\kB \ln{(|\cN|)}
\end{equation}
is the exponential rate at which the number of network configurations grows with the total number of atoms.

\begin{figure}[!t]
  \includegraphics[width=0.45\textwidth]{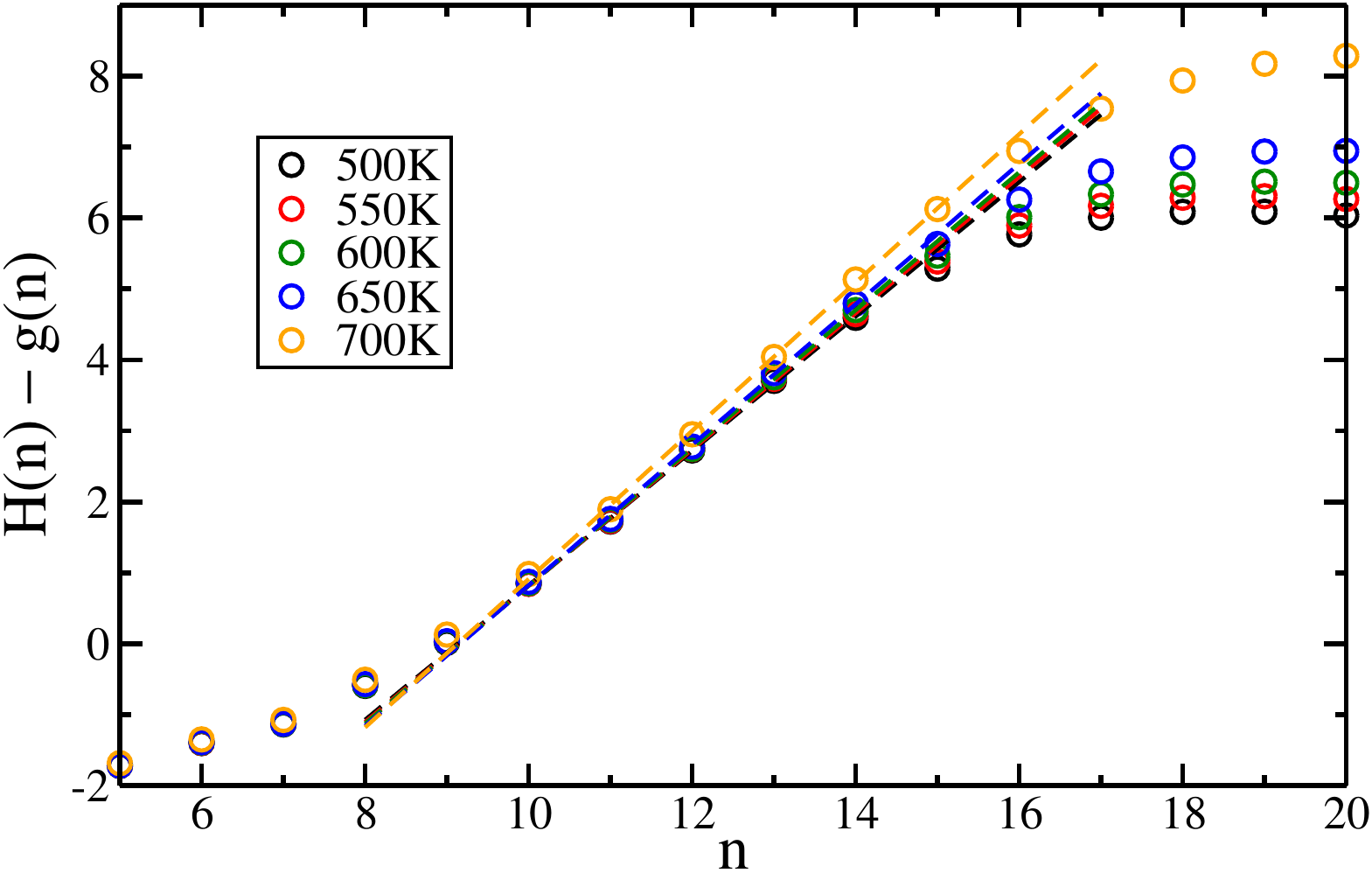}
  \caption{\label{fig:Sconf} $H_c(n)$ vs. $n$ for selected temperatures.}
\end{figure}

For our simulated structure at 600K the vibrational density of states and its associated thermal properties are shown in Fig.~\ref{fig:vDOS}. The DOS and resulting thermodynamic parameters are essentially independent of the simulation temperature, justifying our approximation that $F_{\rm vib}$ is nearly independent of the network configuration. Because of our large 4096-atom cell size we can approximate the integral in Eq.~\eqref{eq:F_vib} with a sum over the non-zero frequencies evaluated at the phonon $\Gamma$-point. We obtain these frequencies by diagonalizing the dynamical matrix calculated from finite differences of total energy within our ML potential.

We obtain the configurational entropy of the covalent bonding network through an information-theoretic method due to Vink and Barkema~\cite{Barkema}. Each network can be considered as a graph consisting of vertices (atoms) and edges (bonds). Subgraphs of the $n$ nearest neighbors are based on each atom and each subgraph (up to irrelevant automorphisms that permute the atoms) is assigned a label $\lambda$. For each number of neighbors $n$ we gather the probability distributions $\{p_n(\lambda)\}$, then compute the Shannon entropy of the distribution,
\begin{equation}
  \label{eq:Shannon}
  H(n) = - \sum_\lambda p_n(\lambda)\ln{p_n(\lambda)}.
\end{equation}
The entropy per atom
\begin{equation}
  \label{eq:S_conf}
  S_{\rm conf} = \Lim{n\to\infty}[H(n+1)-H(n)]
\end{equation}
is the rate at which the diversity of the set of subgraphs grows with each additional vertex. An artifact caused by the sensitivity of graph structure to atomic displacement is resolved by replacing $H(n)$ with $H(n)-(d-1)\ln{n}$. Fig.~\ref{fig:Sconf} shows our results for several temperatures. Further details of these calculations are in the Appendix.

Fig.~\ref{fig:Gibbs} the Gibbs free energies of the LDA and HDL phases, and also the diamond crystalline (cF8) structure). The free energies of LDA and HDL cross in the vicinity of 637K, while their free energies lie above the crystal indicating the transition occurs in the metastable state.

\begin{figure}
  \includegraphics[width=0.45\textwidth]{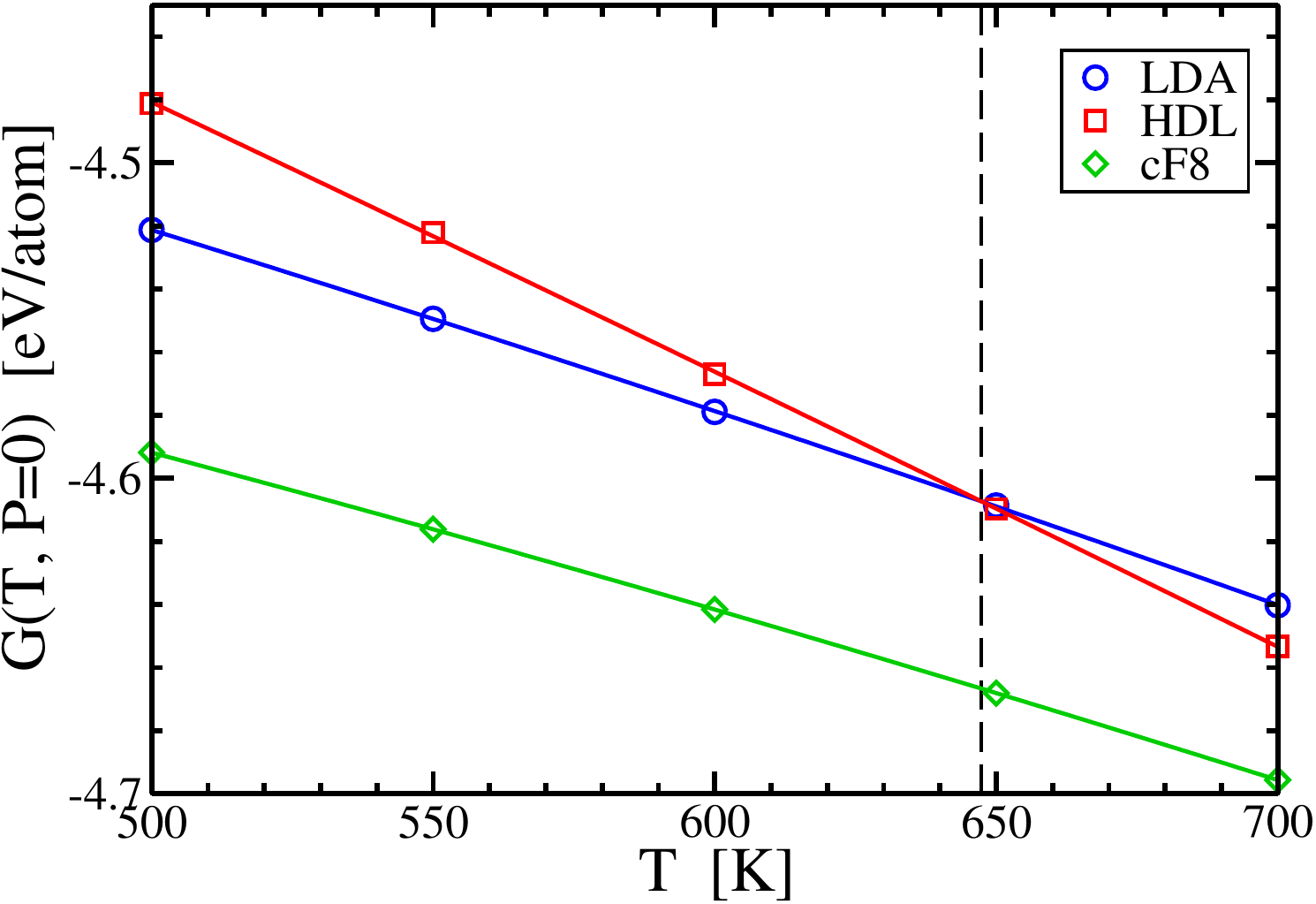}
  \caption{\label{fig:Gibbs} Gibbs free energies of HDL, LDA and cF8 phases}
\end{figure}

Our result suggests the following scenario for explosive recrystallization. Starting with an amorphous thin film of Ge supported on a substrate, a localized heating can transform the LDA state into HDL. This metastable liquid state in turn will rapidly crystalize to form the low temperature ground state structure. Note that the quench rate of $10^{10}$/sec in our hysteresis simulation far exceeds a likely cooling rate in the laboratory.

\bibliography{refs}

\appendix

\renewcommand{\theequation}{\thesection\arabic{equation}}
\setcounter{equation}{0}
\makeatletter
\@addtoreset{equation}{section}
\makeatother
\renewcommand{\thefigure}{\thesection\arabic{figure}}
\renewcommand{\thetable}{\thesection\arabic{table}}

\makeatletter
\@addtoreset{figure}{section}
\@addtoreset{table}{section}
\makeatother



\newpage
\section{End Matter}
\label{app:A}


{\em first principles calculations---}Our molecular dynamics and phonon calculations utilize the VASP code~\cite{VASP-PAW} in the PBE generalized gradient approximation~\cite{PBE} using the supplied projector augmented wave potentials for Ge with valence 4 at the default energy cutoff. We employ a single $k$-point at (\textonequarter\textonequarter\textonequarter) which efficiently samples the Brillouin zone because the cell is large and the structures are symmetric on average~\cite{Baldereschi,Galtsov}. For computational efficiency we employ machine-learning (ML) interatomic potentials~\cite{VASP-ML} trained on data sets of 512 atoms. We trained the potential for the hysteresis loop on a data set that combined both low and high density amorphous structures. Potentials for free energy calculations were trained separately for each amorphous phase and for the diamond structure.

\begin{figure}[b]
  \includegraphics[width=0.45\textwidth]{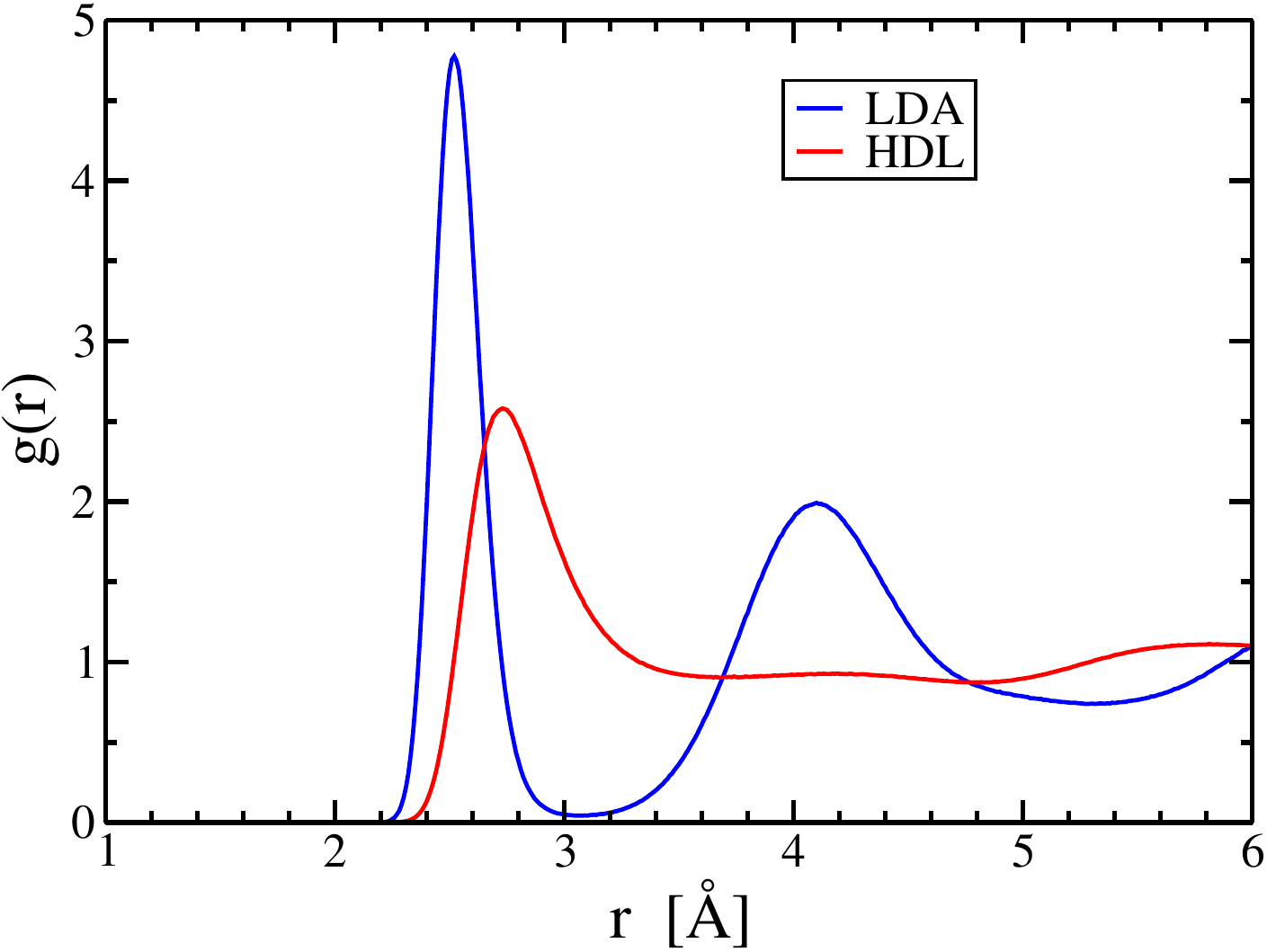}
  \caption{\label{fig:pdf600} Pair distribution functions $g(r)$ at T=600K.}
\end{figure}

\begin{figure}[h]
  \includegraphics[width=0.45\textwidth]{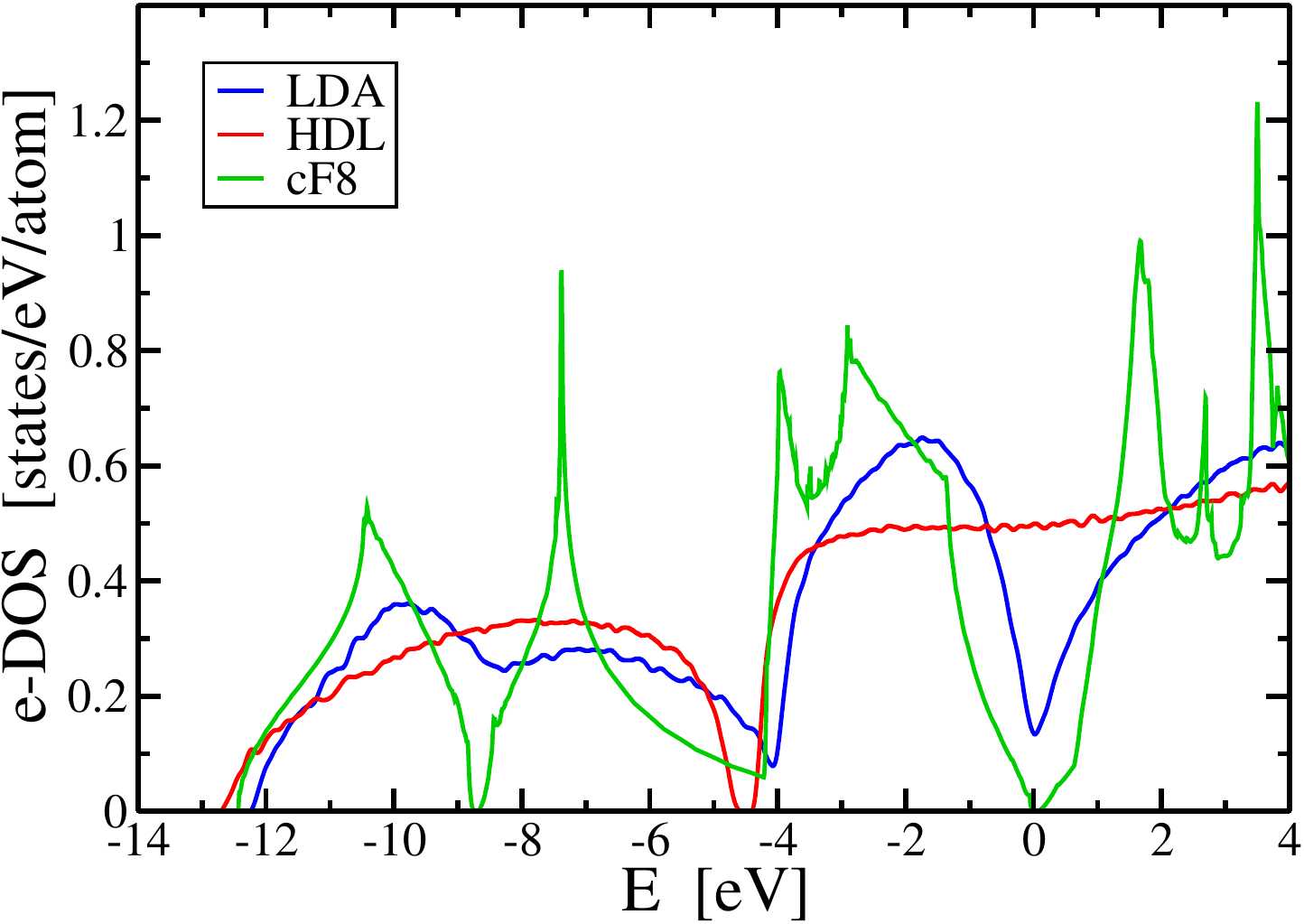}
  \caption{\label{fig:eDOS} e-DOS of low density amorphous and high density liquid structures T=700K, and additionally for the low temperature crystalline state.}
\end{figure}

Molecular dynamics simulations, phonon and electronic density of states, and free energy calculations, were performed on cells of 4096 atoms. The hysteresis loop required $10^7$ time steps of 5fs in the NPT ensemble, for a total simulated time of 10 ns. Each free energy data point employed 10fs time steps in the NVT ensemble and reached simulated times of 4ns for the metal and reached times of 80ns for the LDA network. Equilibrium volumes for HDL were determined separately at each temperature, while they were held constant for LDA, as its thermal expansion is negligible. Representative simulated pair distribution functions of the metal and the $\sp3$ network are shown in Fig.~\ref{fig:pdf600}. Electronic densities of states are given in Fig.~\ref{fig:eDOS}. Table~\ref{tab:F_cF8} provides free energy components of the crystalline state.

{\em supplemental figures and table---}

\begin{figure}[h]
  \includegraphics[width=0.45\textwidth]{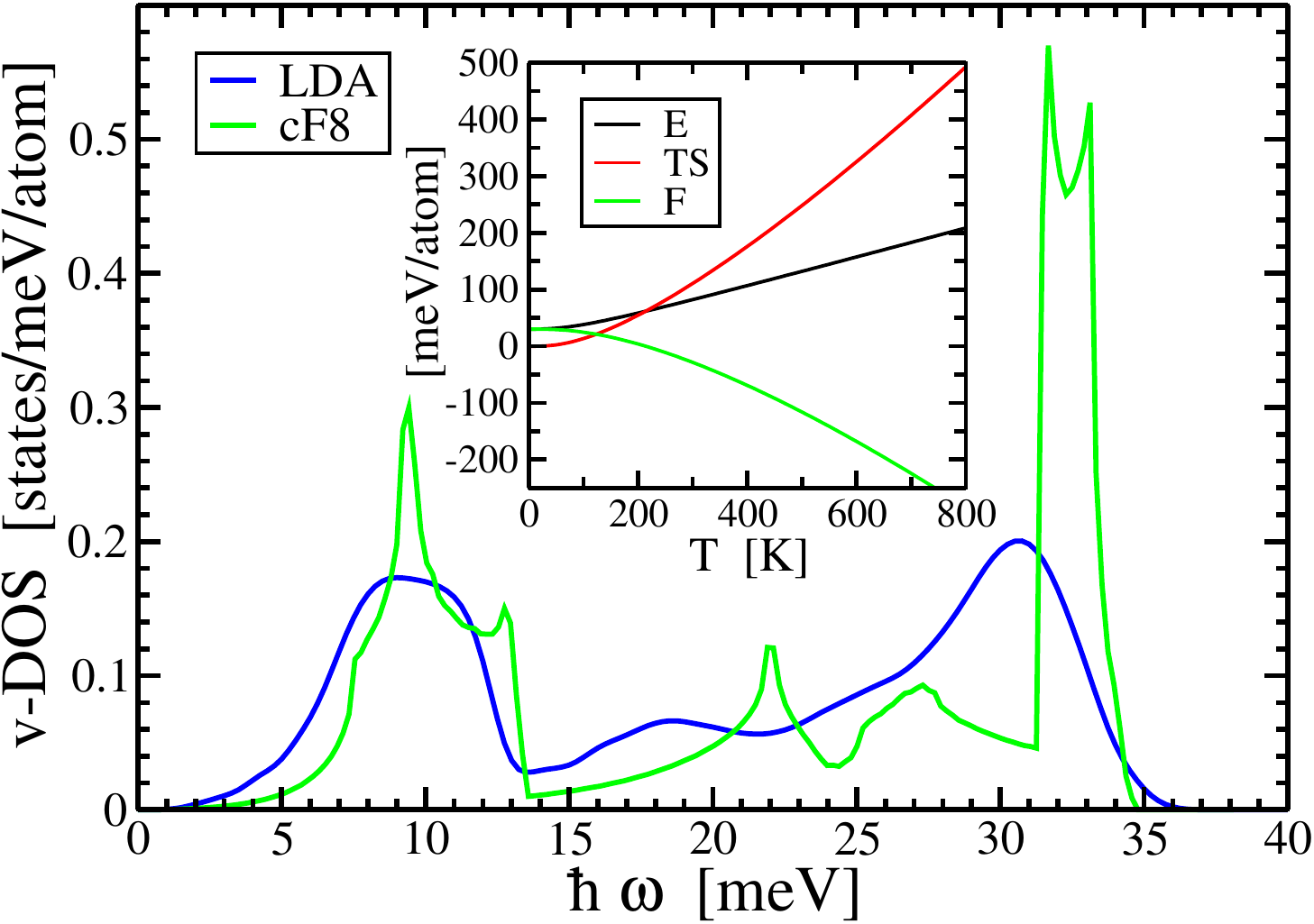}
  \caption{\label{fig:vDOS} Vibrational density of states of low density amorphous germanium compared with its crystalline counterpart. Inset shows vibrational energy, entropy and free energy of the LDA state.}
\end{figure}

\begin{table}[b]
  \begin{tabular}{c|rrrrr}
        T        &    500 &    550 &    600 &    650 &    700\\
  \hline
  $\avg{K}$      &  0.065 &  0.071 &  0.078 &  0.084 &  0.090\\
  $\avg{E_0}$    & -4.488 & -4.488 & -4.487 & -4.487 & -4.487\\
  $\avg{H}$      & -4.423 & -4.416 & -4.410 & -4.403 & -4.396\\
  \hline
  $S_{\rm vib}$  &  5.758 &  6.039 &  6.294 &  6.533 &  6.761\\
  \hline
  $G$            & -4.521 & -4.549 & -4.580 & -4.609 & -4.640
  \end{tabular}
  \caption{\label{tab:F_cF8} Entropy terms (in units of $\kB$/atom), energies (units of eV/atom) and free energies of the Pearson type cF8 crystal. Gibbs free energy $G=H-TS$ with $S=S_{\rm vib}$.}
\end{table}

\begin{figure*}[h]
  \centering
  \includegraphics[trim = 0cm 14cm 0cm 0cm, clip, width=0.95\textwidth]{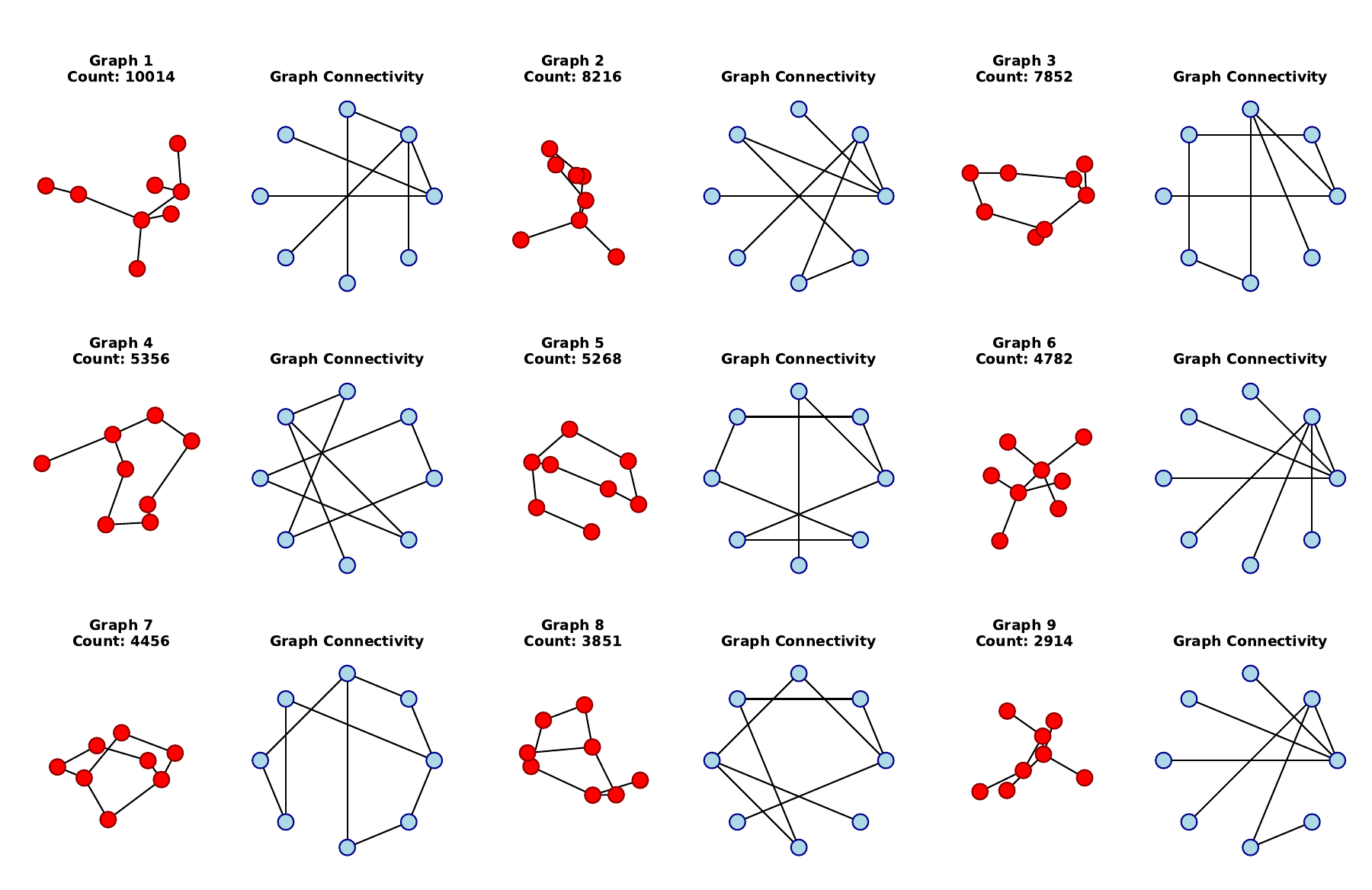}
  \caption{Representative local topological structures, their corresponding graph representations, and the counts of configurations within each equivalence class. The structures are sampled from a single MD frame at 700 K using a total of $10^5$ uniformly distributed random positions. A typical real-space local structure is shown with atoms highlighted in red. The corresponding graph representation, with vertices shown in cyan, indicates the isomorphism class of the local topology.}
  \label{fig:b}
\end{figure*}

{\em Network entropy---}Following the method described in Ref.~\cite{Barkema}, we evaluate the configurational entropy of network-forming $\alpha$-Ge. The nearest-neighbor bond length is determined from the pair distribution function (Fig.~\ref{fig:b}); the first peak, located at $r \approx 3.1$~\AA, corresponds to the first coordination shell.
\begin{figure}[htpb]
  \centering
  \includegraphics[width=.45\textwidth]{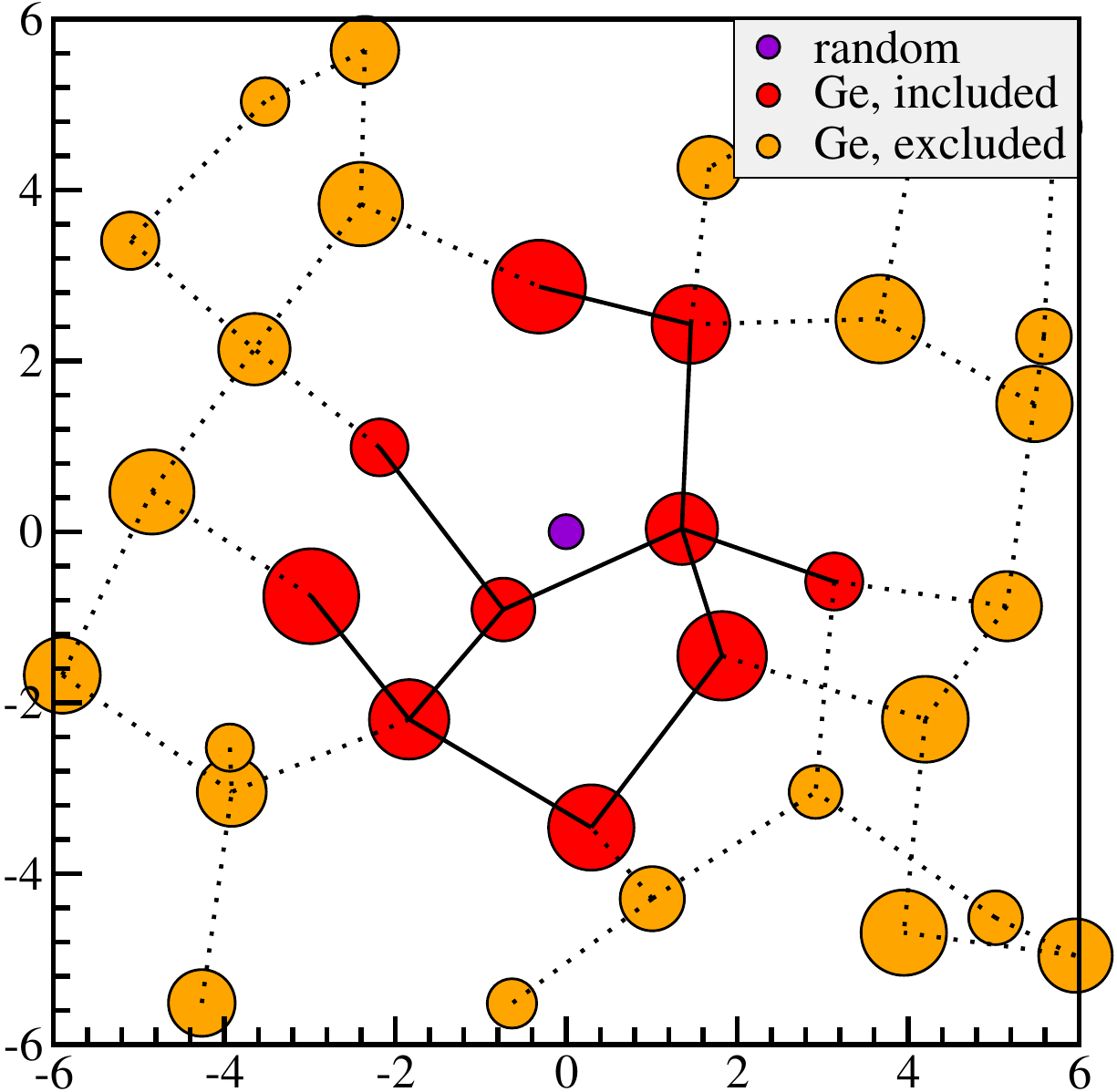}
  \caption{(a) Local topology structure near a randomly selected point.}
  \label{fig:graph}
\end{figure}

We compute the configurational entropy using the topological sampling approach of Ref.~\cite{Barkema}, in which local environments are classified into equivalence classes based on graph isomorphism. Specifically, we sample local topological structures from 2000 MD frames. For each frame, $10^5$ random points are generated uniformly within the simulation cell (see, e.g. Fig.~\ref{fig:graph}. For each sampled point, we identify the closest $n$ atoms, which define a local topology. Each local topology is mapped onto a graph representation, with atoms as vertices and bonds as edges. Graph isomorphism analysis is then performed using pynauty \cite{mckay2014} to classify the graphs into topologically equivalent classes, from which the configurational entropy is computed.

In Fig.~\ref{fig:b}, we present representative local topological structures, their corresponding graph representations, and the counts of configurations within each equivalence class. From these counts, we estimate the probability $p(\lambda)$ of each equivalence class based on its frequency. The total entropy $S(n)$ of $n$-atom clusters is then given by \cite{Barkema},
\begin{eqnarray}
  \label{eq:a}
  H(n) = -\sum_ip(\lambda)\log p(\lambda);\\
  g(n)=(d-1) \log (n);\\
  S(n)/k_{\rm B} = H(n) - g(n);
\end{eqnarray}
where $p(i)$ denotes the estimated probability of equivalence class $\lambda$, and $d$ is the dimensionality of the network.

\end{document}